\documentclass{ws-procs9x6}

\begin{document}

\title{Stripes and Charge Transport Properties of High-$T_c$ Cuprates}

\author{Yoichi Ando}

\address{Central Research Institute of Electric Power Industry,\\
Komae, Tokyo 201-8511, Japan\\ 
E-mail: ando@criepi.denken.or.jp}


\maketitle

\abstracts{ 
Unusual features in the in-plane charge transport in lightly hole-doped
La$_{2-x}$Sr$_x$CuO$_4$ single crystals are described. Notably, both the
in-plane resistivity and the Hall coefficient show a metallic behavior at
moderate temperatures even in the long-range-ordered antiferromagnetic
phase, which obviously violates the Mott-Ioffe-Regel criterion for the
metallic transport and can hardly be understood without employing the
role of charge stripes. Moreover, the mobility of holes in this
``metallic" antiferromagnetic state is found to be virtually the same as
that in optimally-doped crystals, which strongly suggests that the
stripes govern the charge transport in a surprisingly wide doping range
up to optimum doping.}

\section{Introduction}

In high-$T_c$ cuprates such as La$_{2-x}$Sr$_x$CuO$_4$ (LSCO), the
antiferromagnetic (AF) state gives way to high-$T_c$ superconductivity
when a sufficient number of holes are doped into the CuO$_2$ planes. The
AF state of cuprates is therefore a natural starting point to establish
the picture of high-$T_c$ superconductors, but nevertheless their
transport properties have not drawn sufficient attention. It has been
generally believed that the hole motion inevitably frustrates the
antiferromagnetic bonds and thus the doped holes must be strongly
localized until the long-range AF order is destroyed. Indeed, the
variable-range-hopping conductivity has been mostly observed in the AF
state of cuprates,\cite{Kastner,Keimer} which is naturally expected for
the localized holes. As a result, researchers have been discouraged by
the apparent simplicity of this so-called ``antiferromagnetic insulator"
regime.

However, recent measurements in clean, lightly-doped YBa$_2$Cu$_3$O$_y$
(YBCO) crystals have demonstrated\cite{Ando,Lavrov} that the charge
transport in the AF state is full of surprise: the temperature
dependence of the in-plane resistivity $\rho_{ab}$ remains to be
metallic ($\rho_{ab}$ decreases with decreasing temperature) across the
N\'{e}el temperature $T_N$, anomalous features in the magnetoresistance
imply that holes form stripes instead of being homogeneously
distributed, and along the $c$-axis the charge confinement
characteristics are significantly affected by the N\'{e}el ordering.
Motivated by these results\cite{Ando,Lavrov} on YBCO that we obtained in
1999, we have revisited the charge transport in clean single crystals of
LSCO, where studying the lightly-doped regime is much more
straightforward than in other cuprates; the hole doping $p$ in the
CuO$_2$ planes is equal to $x$, the Sr content, and $T_N$ can be readily
determined by susceptibility measurements.\cite{Thio} 

Here we show that, contrary to the common belief, the doped holes in
clean single-crystalline cuprates are surprisingly mobile in a wide
range of temperatures even in the long-range-ordered AF phase. This is
possible when the electron system self-organizes into hole-rich stripes
and hole-poor AF regions to facilitate the motion of charges. We further
show that the hole mobility at moderate temperatures remains virtually
unchanged throughout a wide doping range from the lightly-doped AF
regime (hole doping of 1\%) to the optimally-doped regime (hole doping
of 17\%) where the superconducting transition temperature is maximal.
This strongly suggests that the hole motion is governed by the stripes
all the way up to optimum doping, and thus the high-temperature
superconductivity appears to be a property associated with the stripes.

\begin{figure}[bt]
\epsfxsize=1\columnwidth
\centerline{\epsffile{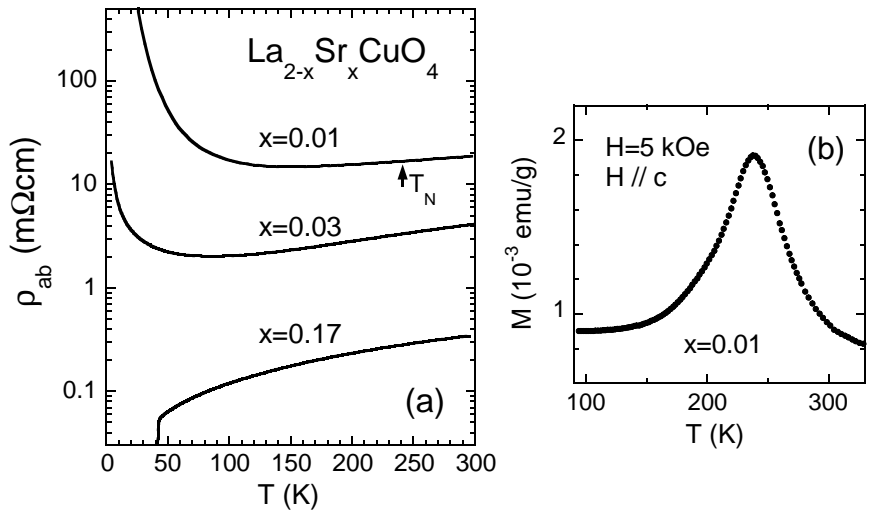}}
\caption{(a) Temperature dependences of 
$\rho_{ab}$ of lightly-doped ($x = 0.01$ and 0.03) and 
optimally-doped ($x = 0.17$) La$_{2-x}$Sr$_x$CuO$_4$ single crystals. 
(b) Magnetization of a large La$_{1.99}$Sr$_{0.01}$CuO$_4$ 
single crystal from which the samples for $\rho_{ab}$ measurements 
were cut; the peak in $M(T)$ corresponds to the N\'{e}el temperature.}
\label{fig1}
\end{figure}

\section{Experimental}

The clean single crystals of LSCO are grown by the 
traveling-solvent floating-zone (TSFZ) technique\cite{Komiya} and 
are carefully annealed to remove excess oxygen, 
which ensures that the hole doping is exactly equal to $x$.  
The in-plane resistivity $\rho_{ab}$ and the Hall coefficient 
$R_H$ are measured using a standard ac six-probe method. 
The Hall effect measurements are done by sweeping the 
magnetic field to $\pm$14 T at fixed temperatures stabilized 
within $\sim$1 mK accuracy.\cite{Ando}
The Hall coefficients are always determined by fitting the 
$H$-linear Hall voltage in the range of $\pm14$ T, 
which is obtained after subtracting the 
magnetic-field-symmetrical magnetoresistance component 
caused by small misalignment of the voltage contacts.

\section{Results}

\begin{figure}[bt]
\epsfxsize=80mm
\centerline{\epsffile{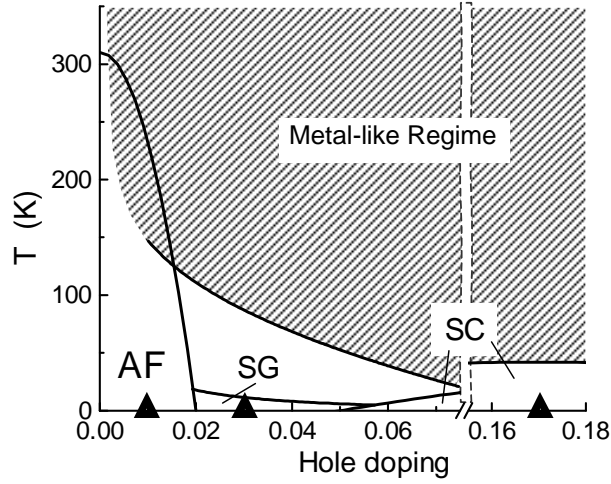}}
\caption{The antiferromagnetic (AF), spin-glass (SG) and 
superconducting (SC) regions on the phase diagram of LSCO; 
representative doping levels chosen for this article are indicated 
by triangles. The hatched region illustrates where $\rho_{ab}$ 
shows the metal-like behavior ($d\rho_{ab}/dT > 0$).}
\label{fig2}
\end{figure}

Figure 1(a) shows the temperature dependences of $\rho_{ab}$ for LSCO
crystals which represent three doping regimes\cite{Kastner,Niedermayer}
on the phase diagram (Fig. 2): antiferromagnetic [the sample with $x =
0.01$ has $T_N \simeq 240$ K according to the magnetization
data shown in Fig. 1(b)], spin glass ($x = 0.03$), and
optimally-doped superconductor ($x = 0.17$). (More complete data sets
can be found in our recent papers.\cite{Komiya,mobility}) One may notice
that, while the magnitude of the resistivity significantly increases
with decreasing doping, the temperature dependence at $T > 150$ K does
not change much; in particular, in the sample with $x = 0.01$,
$\rho_{ab}$ keeps its metallic behavior well below $T_N$. This
observation in the lightly-doped LSCO crystal clearly invalidates the
long-standing notion that the metal-like behavior of $\rho_{ab}(T)$ in
cuprates may appear only as soon as the long-range AF order is
destroyed. 

\begin{figure}[t!]
\epsfxsize=0.7\columnwidth
\centerline{\epsffile{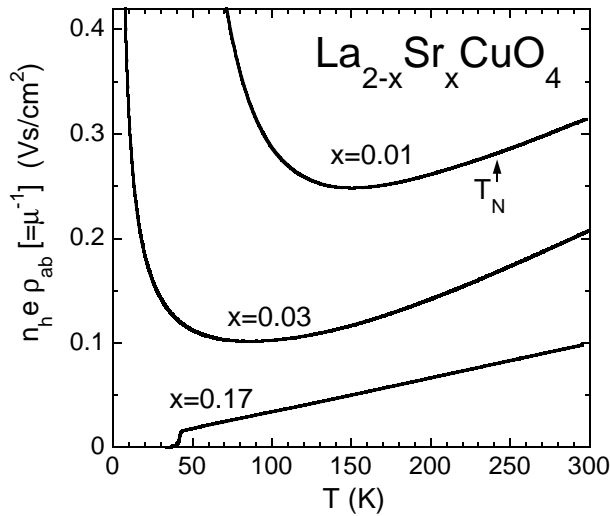}}
\caption{Temperature dependences of the normalized resistivity 
$n_{h}e\rho_{ab}$ of LSCO crystals, where $n_h = 2x/V$ is the 
nominal hole density.  
Note that $n_{h}e\rho_{ab}$ is essentially an inverse mobility 
$\mu^{-1}$ of doped holes.}
\label{fig3}
\end{figure}

To examine whether the hole mobility actually depends on the magnetic
state as crucially as has been expected, in Fig. 3 we normalize
$\rho_{ab}$ by the nominal hole concentration $n_h$, which is given by
$2x/V$ [unit cell $V$ ($\simeq 3.8\times 3.8\times 13.2$ \AA$^3$)
contains two CuO$_2$ planes]. The product $n_{h}e\rho_{ab}$ would mean
just inverse hole mobility $\mu^{-1}$ if we assume the number of mobile
holes to be always given by $x$. Apparently, the slope and magnitude of
$n_{h}e\rho_{ab}$ at moderate temperatures are very similar, suggesting
that the transport is governed by essentially the same mechanism for all
three doping regimes; in particular, the magnitudes of the hole mobility
at room temperature differ by only a factor of three between $x$ = 0.01
and 0.17, demonstrating that the hole mobility remains virtually
unchanged in a surprisingly wide range of doping. We note that the
magnitude of the hole mobility in LSCO (order of 10 cm$^2$/Vs at 300 K)
is almost the same as that in YBCO;\cite{mobility} this suggests that
the hole mobility in the CuO$_2$ planes is essentially universal among
the cuprates. Interestingly, typical metals (such as iron or lead) show
similar values of carrier mobility, $(ne\rho)^{-1}$, at room
temperature.\cite{mobility}

\begin{figure}[t!]
\epsfxsize=0.7\columnwidth
\centerline{\epsffile{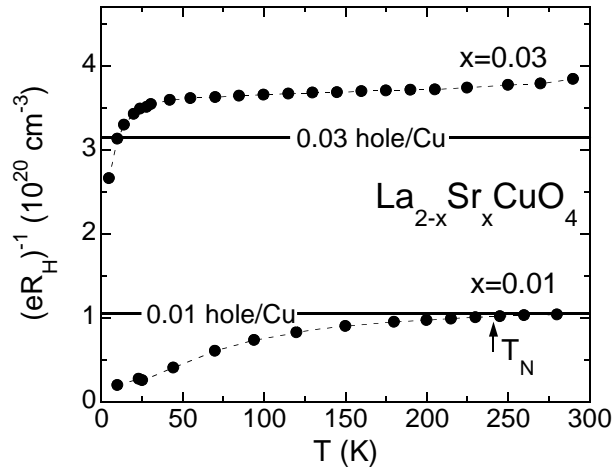}}
\caption{The apparent hole density of carriers $n = (eR_H)^{-1}$ 
for the two lightly-doped LSCO crystals; 
solid lines indicate the nominal value $n_h$.}
\label{fig4}
\end{figure}

The region that is characterized by the metallic transport behavior
($d\rho_{ab}/dT > 0$) is depicted in the phase diagram (Fig. 2);
evidently, it extends widely in the phase diagram and essentially
ignores the changes in the magnetic properties. It is worth noting that
the normal-state resistivity in superconducting LSCO was 
studied\cite{logT,Boebinger} by suppressing superconductivity with 60-T
magnetic fields and an increase in $\rho_{ab}$ at low temperature was
observed up to optimum doping; thus, the high mobility of holes at
moderate temperature and localization at low temperature appear to be
essentially unchanged in the normal state in the whole underdoped
region, all the way from $x$ = 0.01 to 0.15.

Another evidence for unexpected metallic charge transport in the AF
cuprates can been found in the Hall coefficient $R_H$. The apparent hole
density $n = (eR_{H})^{-1}$ obtained for the LSCO samples with $x =
0.01$ and 0.03 (Fig. 4) is essentially temperature independent in the
temperature range where the metallic behavior of $\rho_{ab}(T)$ is
observed, which is exactly the behavior that ordinary metals show.
Moreover, $n$ agrees well with the nominal hole concentration $n_h =
2x/V$ at $x = 0.01$, which means that all the doped holes are moving and
contributing to the Hall effect even in the long-range-ordered AF state
down to not-so-low temperatures until disorder causes the holes to
localize. For higher doping, the ratio $n/n_h$ exceeds unity and reaches
a value of $\sim$3 at optimum doping. 

\section{Discussions}

\subsection{Unusual Metallic Transport}

It is useful to note that the absolute value of $\rho_{ab}$ for $x$ =
0.01 is as large as 19 m$\Omega$cm at 300 K. If we calculate the
$k_{F}l$ value ($k_{F}$ is the Fermi wave number and $l$ is the mean
free path) using the formula $hc_0/\rho_{ab}e^2$ ($c_0$ is the
interlayer distance), which implicitly assumes a {\it uniform} 2D electron
system and the Luttinger's theorem, the $k_{F}l$ value for $x$ = 0.01
would be only 0.1; this strongly violates the Mott-Ioffe-Regel limit for
metallic transport, and thus the conventional wisdom says that the
band-like metallic transport is impossible for $x$ = 0.01. In other
words, the metallic transport in the slightly hole-doped LSCO is a
strong manifestation of the ``bad metal" behavior.\cite{BadMetal}

Very recent angle-resolved photoemission spectroscopy (ARPES)
measurements of lightly-doped LSCO crystals have found\cite{Yoshida}
that ``Fermi arcs" develop at the zone-diagonal directions in the
$\mathbf{k}$-space, on which metallic quasiparticles are observed. These
Fermi arcs are different from the small Hall pockets and apparently
violate the Luttinger's theorem, because the Fermi surface is partially
destroyed and thus the enclosed {\it area} is not well-defined. Therefore,
at least phenomenologically, such violation of the Luttinger's theorem by
the Fermi arcs allows the system to have a small effective carrier
number and a ``large" $k_{F}$ value at the same time, which enables the
metallic transport to be realized in the lightly hole-doped regime.
(Thus, the $k_{F}l$ value estimated under the assumption of a uniform 2D
system is obviously erroneous.)

\subsection{Difficulty of Metallic Transport in the Antiferromagnetic State}

How can such an unusual metallic transport and the relatively high
mobility of doped holes be possible in the long-range-ordered AF phase?
It has been known for a long time that a single hole doped into a
two-dimensional square antiferromagnet should have a very low mobility
because of the large magnetic energy cost of the spin bonds broken by
the hole motion, although quantum effects allow the hole to
propagate.\cite{Dagotto,review} Despite this common knowledge, our
resistivity and the Hall coefficient data demonstrate that the doped
holes in the AF state can have the mobility nearly as high as that at
optimum doping, which means that the holes manage to move without paying
the penalty for frustrating AF bonds. This striking contrariety is not
restricted to the simple one-band model implicitly hypothesized in the
above argument. Whatever the transport mechanism is, the doped holes
should have an extremely strong coupling to the AF background; otherwise
such a small amount of holes as 2\% would not be able to destroy the AF
state.\cite{Kastner} At the same time, this strong coupling tends to
localize the holes arbitrarily distributed in the AF background, since
the spin distortion created by a hole in the rigid N\'{e}el state
destroys the translational symmetry. Therefore, the unusually metallic
charge transport in the AF phase requires a novel mechanism to be
realized in the lightly-doped cuprates.

\subsection{Role of Stripes}

To the best of our knowledge, the only possibility for the metal-like
conductivity to survive under the strong coupling of holes with the
magnetic order is when the holes and spins form a superstructure which
restores the translational symmetry. A well-known example is the striped
structure,\cite{review,Kivelson,Zaanen} where the energy cost for the
distortion of the spin lattice is paid upon the stripe formation and
then the holes can propagate along the stripes without losing their
kinetic energy. In fact, the striped structure has been already
established\cite{Tranquada} for La$_{2-x-y}$Nd$_y$Sr$_x$CuO$_4$, and
there is now growing evidence for the existence of stripes in other
hole-doped cuprates,\cite{Ando,Yamada,Mook,Hunt,anisotropy} the case
being particularly strong for LSCO and YBCO in the lightly-doped region.
Moreover, the mesoscopic phase segregation into the metallic paths
(charge stripes) and the insulating domains (AF regions) offers a
natural explanation about why the apparent $k_{F}l$ value can be so
small in the regime where metallic transport is observed.\cite{mobility}
Existence of such charged magnetic-domain boundaries are actually
indicated by our recent in-plane anisotropy measurements of the magnetic
susceptibility of lightly-doped LSCO.\cite{sus}

One might wonder about the nature of the Hall effect when the
conductivity occurs through the quasi-one-dimensional (1D) stripes.
Indeed, it was shown that the Hall effect tends to disappear in
La$_{1.4-x}$Nd$_{0.6}$Sr$_x$CuO$_4$ (LNSCO) upon the transition into the
static stripe phase.\cite{Noda} Against our intuition, however, the
quasi-1D motion itself does {\it not} necessarily drive the Hall
coefficient to zero. The quasi-1D confinement dramatically suppresses
the transverse (Hall) {\it current} induced by the magnetic field, but
the same large transverse resistivity restores the finite Hall {\it
voltage}, because $R_H \sim \sigma_{xy}/\sigma_{yy}\sigma_{xx}$. For the
same reason, for instance, the well-known charge confinement in the
CuO$_2$ planes in cuprates does not prevent generation of the Hall
voltage along the $c$-axis ($H \parallel ab$).\cite{Harris} Therefore,
the Hall-effect anomaly in LNSCO must be caused by some more elaborate
mechanism rather than simply due to the quasi-1D nature of the
transport. One possibility is that the anomaly in LNSCO is due to the
peculiar arrangement of stripes which alter their direction from one
CuO$_2$ plane to another and thereby keeping $\sigma_{yy}$ from
vanishing; on the other hand, the unidirectional stripes\cite{Matsuda}
in pure LSCO would naturally keep the Hall coefficient unchanged, and
thus the apparently contrasting behavior of the Hall effect in
lightly-doped LSCO and LNSCO can be compatible with the existence of the
stripes in both systems. Another possible source of difference between
the two systems is the particle-hole symmetry inside the stripes: It has
been proposed that the vanishing Hall coefficient in LNSCO is
essentially due to the particle-hole symmetry realized by the 1/4-filled
nature of the stripes near the 1/8 doping;\cite{EmeryHall,Prelovsek} if,
on the other hand, the stripes at small $x$ values are not exactly 1/4
filled, it is natural to observe non-vanishing Hall coefficient in LSCO,
in the context of these theories.\cite{EmeryHall,Prelovsek} Also, it is
possible that the finite Hall resistivity in LSCO is caused the
transverse sliding of the stripe as a whole; in fact, very recent
optical conductivity measurements of lightly-doped LSCO have concluded
that the sliding degrees of freedom are important for the realization of
the metallic transport in this system.\cite{Dumm}

From the above discussion, it is clear that the metallic in-plane charge
transport we observe in the AF state is most likely governed by the
charge stripes. Given the fact that the hole mobility at moderate
temperatures is surprisingly insensitive to the hole doping all the way
up to optimum doping, it is tempting to conclude that the charge
transport in cuprates that show the maximal $T_c$ is also governed by
the stripes. Recent STM studies of optimally-doped
Bi$_2$Sr$_2$CaCu$_2$O$_{8+\delta}$ compounds, where periodic spacial
modulations of the local density of state are
observed,\cite{Kapitulnik,Davis} also seem to support this conclusion.
The implication of such a conclusion on our understanding of the
high-$T_c$ superconductivity is rather significant. Since the ordered
static stripes are known to {\it kill} superconductivity, it must be the
fluctuating nature of the stripes that facilitate the superconductivity
at such high temperatures. There are already some theoretical proposals
to explain the high-$T_c$ superconductivity on the basis of the
fluctuating stripes\cite{review,Emery,Carlson} or charge
fluctuations.\cite{Castellani} The system we are dealing with may indeed
be the ``electronic liquid crystals",\cite{Kivelson} which are
quantum-fluctuating charge stripe states; our recent studies of the
in-plane resistivity anisotropy of lightly-doped cuprates have
found\cite{anisotropy} that the resistivity is smaller along the stripe
direction but the magnitude of the anisotropy is strongly dependent on
temperature, which suggests a crossover between different electronic
liquid crystal phases occurring in the cuprates, and the low-temperature
phase appears to be an electron nematics.\cite{Fradkin} Clearly, more
experiments are needed to fully understand such a new state of matter,
and to finally elucidate the mechanism of the high-$T_c$
superconductivity.

\section{Summary}

It is shown that the doped holes in cuprates are surprisingly mobile in
the long-range-ordered antiferromagnetic state at moderate temperatures,
which is evidenced both by the metallic $\rho_{ab}(T)$ behavior and by
the almost temperature-independent $R_H(T)$. It is emphasized that the
{\it mobility} of the doped holes at moderate temperatures is virtually
unchanged from the lightly hole-doped antiferromagnetic compositions
(where the dominance of the stripes is very likely) to the
optimally-doped superconducting composition, which implies that the
charge transport even at optimum doping is essentially governed by the
stripes.

\section*{Acknowledgments}
This work was done in collaboration with A. N. Lavrov, S. Komiya, X. F.
Sun, and K. Segawa.  Stimulating discussions with S. A. Kivelson are
greatly acknowledged.  We also thank D. N. Basov, A. Fujimori, and 
J. M. Tranquada for collaborations and helpful discussions.

\end{document}